\documentclass[fleqn,usenatbib]{mnras}
\usepackage{newtxtext,newtxmath}
\usepackage[T1]{fontenc}

\DeclareRobustCommand{\VAN}[3]{#2}
\let\VANthebibliography\thebibliography
\def\thebibliography{\DeclareRobustCommand{\VAN}[3]{##3}\VANthebibliography}

\usepackage{graphicx}
\usepackage[export]{adjustbox}
\usepackage{amsmath}
\usepackage{booktabs}
\usepackage{threeparttable}
\usepackage{placeins}
\usepackage{caption}
\usepackage{threeparttable}

\usepackage{xcolor}
\definecolor{lb}{RGB}{44, 139, 183}

\usepackage{color,soul}
\usepackage[normalem]{ulem}
\title[Gamma-ray structure around the Coma cluster]{Detailed study of extended gamma-ray morphology in the vicinity of the Coma cluster with \textit{Fermi}-LAT}

\author[Baghmanyan et al. 2021]{
Vardan Baghmanyan,$^{1,2}$\thanks{E-mail:vardan.baghmanyan@ifj.edu.pl}
Davit Zargaryan,$^{3,4}$
Felix Aharonian,$^{3,4,2}$
Ruizhi Yang,$^{5}$
\newauthor{
Sabrina Casanova$^{1,2}$
and Jonathan Mackey $^{3,4}$
}
\\
$^{1}$Institute of Nuclear Physics PAN, Radzikowskiego 152, 31-342 Krak\'ow, Poland \\
$^{2}$Max-Planck-Institut f\"ur Kernphysik, P.O. Box 103980, 69029 Heidelberg, Germany \\
$^{3}$Dublin Institute for Advanced Studies, 31 Fitzwilliam Place, Dublin 2, Ireland \\
$^{4}$Centre for AstroParticle Physics and Astrophysics, DIAS Dunsink Observatory, Dunsink Lane, Dublin 15, Ireland \\
$^{5}$Department of Astronomy, School of Physical Sciences,University of Science and Technology of China, Hefei, Anhui 230026, China \\
}

\date{Accepted XXX. Received YYY; in original form ZZZ}

\pubyear{2021}

\begin{document}
\label{firstpage}
\pagerange{\pageref{firstpage}--\pageref{lastpage}}
\maketitle

\begin{abstract}

\noindent Galaxy clusters can be sources of high-energy (HE) $\gamma$-ray radiation, due to the efficient acceleration of particles exceeding EeV energies. At present, though, the only candidate for emitting HE $\gamma$-rays is the Coma cluster, towards which an excess of $\gamma$-ray emission has been detected by the Fermi Large Area Telescope (LAT). Using $\mathrm{\sim12.3}$ years of \textit{Fermi}-LAT data, we explored the region of the Coma cluster between energies 100 MeV and 1 TeV by detailed spectral and morphological analysis. In the region of the Coma cluster, we detected diffuse gamma-ray emission of energies between 100 MeV and 1 TeV with a 5.4$\sigma$ extension significance and a 68\% containment radius of $0.82^{+0.10}_{-0.05}$ degrees derived with a 2D homogeneous disk model. The corresponding gamma-ray spectrum extends up to $\sim50$ GeV, with a power-law index of $\mathrm{\Gamma=2.23\pm0.11}$ and flux of $\mathrm{(3.84\pm0.67)\times10^{-12}\,erg\,cm^{-2}\,s^{-1}}$. Using energy arguments we show that point-like sources such as radiogalaxies and star-forming galaxies are unlikely to explain the emission, and more likely, the emission is produced in the Coma cluster. Besides, we also identified three point-like sources in the region. However, due to the limited statistics of the detection, we could neither exclude nor conclude that the total extended emission is contributed to by these three-point like sources.

\end{abstract}


\section{Introduction}

\begin{figure*}
 \centering
\includegraphics[trim={0 0.75cm 0 0}, clip,width=\textwidth]{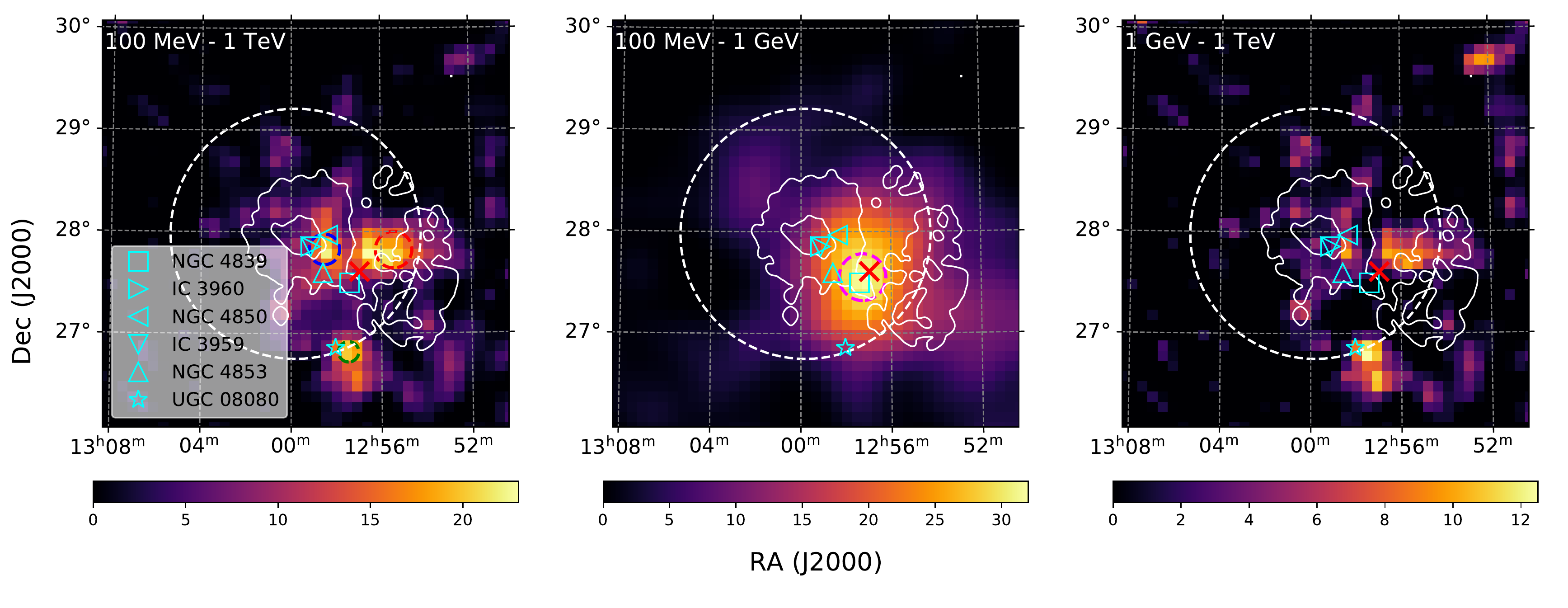}
\caption{TS map of a $\mathrm{4^{\circ} \times 4^{\circ}}$ region centred on the Coma cluster (left: the whole energy range, middle: 100 MeV - 1 TeV, right: 1 GeV - 1 TeV) generated with \textit{tsmap} tool assuming a power-law point source with $\Gamma=2$. The dashed white circle represents the virial region of the Coma cluster ($\theta_{200} = 1.23^{\circ}$). The solid white-colored contours express the radio measurements of the halo and the relic of the Coma cluster with the Westerbork Synthesis Telescope (WSRT) telescope at 352 MHz \citep{2011MNRAS.412....2B}. The cyan-colored markers represent different types of AGNs in the region with NASA/IPAC Extragalactic Data (NED) positions, and the red marker shows the location of the 4FGL J1256.9+2736 source from the \textit{Fermi}-LAT 4FGL-DR2 catalog.
The blue, red, and green colored dashed circles in the left panel and magenta dashed circle in the middle panel correspond to the 68\% containment regions of point-like sources around $\mathrm{p_1}$, $\mathrm{p_2}$, $\mathrm{p_3}$, and $\mathrm{p_{low}}$, respectively.} 
\label{fig:fig1}
\end{figure*}

Clusters of Galaxies are the largest gravitationally bound astrophysical structures in the Universe. They are essential cosmological laboratories for studying a variety of astronomical phenomena, e.g., star formation, the interaction of galaxies,  gravity, particle acceleration, etc. \citep{1996SSRv...75..279V}.
The linear scale of galaxy clusters can extend several Mpc (angular scale of a few degrees). They consist of sometimes thousands of galaxies surrounded by hot and diffuse intracluster gas.
Radio observations of clusters of galaxies have revealed the existence of diffuse structures with a regular and irregular shape, so-called radio halo and relic, with non-thermal spectra implying significant energy density in accelerated particles \citep{Radio1993ApJ...406..399G, 2000NewA....5..335G}. 
The magnetic field strength within the cluster is around a few $\mu$G \citep{2002ARA&A..40..319C, 2004IJMPD..13.1549G, 2010A&A...513A..30B, 2018SSRv..214..122D}, which leads to efficient diffusion of cosmic rays (CRs). Even so, according to \citet{1984ARA&A..22..425H}, galaxy clusters can retain accelerated particles for cosmologically long periods because of their large sizes (up to a few Mpc in diameter). Therefore, CRs can accumulate within a cluster and produce detectable HE $\gamma$-ray emission through collision with ambient matter \citep{1996SSRv...75..279V, 1997ApJ...487..529B}.

A number of astrophysical sources can be responsible for the production and the acceleration of CRs in galaxy clusters.
Supernova-driven galactic winds are considered as potential candidates \citep{1996SSRv...75..279V}, and studies have shown that AGNs located in galaxy clusters could significantly contribute to CRs \citep{1997ApJ...477..560E,2002MNRAS.332..215A, 2007MNRAS.382..466H}.
Furthermore, large-scale intergalactic shocks, generated by the accretion and merger processes \citep{1998APh.....9..227C,2003ApJ...593..599R,2003ApJ...583..695G,2004APh....20..579G} and turbulent re-acceleration \citep{2005MNRAS.363.1173B} have been proposed as acceleration sites.

Studying clusters of galaxies in the $\gamma$-ray band is challenging but can provide important constraints on the efficiency of particle acceleration, magnetic confinement, and radiation processes of CRs \citep{2007IJMPA..22..681B}.
The accelerated CRs can produce HE $\gamma$-rays through leptonic and hadronic radiative processes.
In the hadronic scenario, the inelastic collision of accelerated CR protons with the thermal nuclei in the intra-cluster medium could produce $\gamma$-rays via pion decay \citep{1980ApJ...239L..93D}.
For the leptonic scenario, ultra-relativistic electrons up-scatter target photon fields (e.g., cosmic microwave background (CMB), infrared background) to HE $\gamma$-rays \citep{2000ApJ...535...45A,2003APh....19..679G}.
Alternatively, $\gamma$-ray production may be either from inverse Compton (IC) scattering by secondary electrons in Bethe-Heitler processes, when relativistic protons interact with CMB photon field \citep{2005ApJ...628L...9I, 2008PhRvD..78c4013K}.
Another possibility is synchrotron radiation of ultra-relativistic electrons, which originate from the interactions of VHE $\gamma$-rays with diffuse extragalactic background radiation field \citep{2004A&A...417..391T}.   
Hadronic $\gamma$-ray emission is a more likely radiation mechanism in galaxy clusters than the leptonic one because accelerated HE electrons have relatively short lifetimes (due to energy losses) and do not accumulate to the same extent as protons.
As a result, their spectrum can be characterized by an early cut-off below the HE regime \citep{2000ApJ...535...45A}.

\begin{table*}
\centering
\caption{The results of the morphological analysis of the Coma cluster region with the corresponding statistical errors. The second to fifth columns show the localised positions, position uncertainties at 68\% ($\mathrm{R_{68}^{pos}}$), and TS values of all morphological components. The 68\% containment radii ($\mathrm{R_{68}}$) and extension TS values ($\mathrm{TS_{ext}}$) of the models used in the extension analysis are given in the sixth and seventh columns, respectively.}
\label{tab:table1}
\begin{tabular}{lcccccc}
    \hline  
    Spatial model & $\mathrm{RA_{J2000}}$ & $\mathrm{Dec_{J2000}}$ & $\mathrm{R_{68}^{pos}}$ & TS & $\mathrm{R_{68}}$ & $\mathrm{TS_{ext}}$  \vspace{1mm}\\
    & [deg]& [deg] & [deg] &  & [deg] &\\
    \hline
    \textbf{RadialDisk}  & $194.14\pm0.14$ & $27.38\pm0.13$ & 0.19 & 51.6 & $0.82^{+0.10}_{-0.05}$ & 29.3\vspace{1mm}\\
    \textbf{RadialGaussian} & $194.27\pm0.17$ & $27.56\pm0.17$ & 0.26 & 50.2 & $0.91^{+0.18}_{-0.16}$ & 9.8\vspace{1mm}\\
    \textbf{$\boldsymbol{\mathrm{p_1+p_2+p_3}}$} \\
    $\mathrm{p_1}$  & $194.63\pm0.08$ & $27.83\pm0.13$ & 0.15 & 17.2 & - & -\\
    $\mathrm{p_2}$  & $193.86\pm0.15$ & $27.82\pm0.10$ & 0.18 & 16.4 & - & -\\
    $\mathrm{p_3}$  & $194.37\pm0.06$ & $26.82\pm0.07$ & 0.10 & 15.4 & - & -\\
    \hline
\end{tabular}
\end{table*}

\begin{table*}
\caption{A summary of spectral analysis results. The $\mathrm{log(\mathcal{L})_{max}}$ values, listed in the seventh column, are returned from the binned maximum-likelihood analysis between 100 MeV and 1 TeV range. $\mathrm{\Delta AIC}$ and $\mathrm{\Delta N_{dof}}$ values in the eighth and ninth columns refer to the AIC values and degrees of freedom relative to the uniform disk model, respectively.}
\centering
\label{tab:table2}
\begin{tabular}{lcccccccc}
    \hline
    Spatial model & Photon flux & Energy flux & Index & TS & $\mathrm{N_{pred}}$ & $\mathrm{log(\mathcal{L})_{max}}$ & $\mathrm{\Delta AIC}$ & $\mathrm{\Delta N_{dof}}$ \vspace{1mm}\\
  & $\mathrm{[10^{-9} \times ph \,\, cm^{-2} \, s^{-1}]}$ & $\mathrm{[10^{-12} \times erg \,\, cm^{-2} \, s^{-1}]}$ & & & & & \\
    \hline
    \textbf{$\boldsymbol{\mathrm{Disk}}$} & $5.11\pm1.15$ & $3.84\pm0.67$ & $2.23\pm0.11$ & 51.6 & 863.2 & -313141.3 & - &\vspace{1mm}\\
    \textbf{$\boldsymbol{\mathrm{4FGL J1256.9+2736}}$} & $4.91\pm1.22$ &  $1.81\pm0.36$ & $2.77\pm0.20$ & 29.8 & 685.8 & -313151.9 & 21.2 & 0\vspace{1mm}\\ 
    \textbf{$\boldsymbol{\mathrm{p_1+p_2+p_3}}$} &  &  &  &  &  & -313135.0 & -4.7 & 4\vspace{1mm}\\
    $\mathrm{p_1}$ & $2.12\pm1.16$ & $1.05\pm0.37$ & $2.47\pm0.23$ & 17.2 & 330.8  &  &  & \\
    $\mathrm{p_2}$ & $2.54\pm1.26$ &  $1.16\pm0.39$ & $2.53\pm0.24$ & 16.4 & 388.0 &  &  & \\
    $\mathrm{p_3}$ & $0.85\pm0.59$ &  $0.93\pm0.35$ & $2.08\pm0.24$ & 15.4 & 151.7 &  &  & \vspace{1mm}\\
    \textbf{$\boldsymbol{\mathrm{Disk+p_1+p_2+p_3}}$}  &  &  &  &  &  & -313131.7 & -7.2 & 6\\
    $\mathrm{Disk}$ & $1.42\pm1.26$ &  $1.54\pm1.06$ & $2.09\pm0.26$ & 5.6 & 250.2 &  &  &\\
    $\mathrm{p_1}$ & $1.62\pm1.13$ &  $0.82\pm0.37$ & $2.44\pm0.28$ & 11.1 & 249.7 &  &  & \\
    $\mathrm{p_2}$ & $1.96\pm1.26$ &  $0.86\pm0.43$ & $2.56\pm0.32$ & 8.8 & 289.3 &  &  & \\
    $\mathrm{p_3}$ & $0.51\pm0.50$ &  $0.75\pm0.39$ & $1.99\pm0.30$ & 9.4 & 91.5 &  &  & \\
    \hline
\end{tabular}
\end{table*}
%\begin{threeparttable}
%\end{threeparttable}

The Coma galaxy cluster (Abell 1656) is considered a unique astronomical laboratory because of its proximity ($\mathrm{z=0.023}$; \citealt{redshift1991ApJS...77..363S}) and high total mass ($\mathrm{M_{500}\simeq6\times10^{14}\,M_\odot}$; \citealt{2013A&A...554A.140P}). It has been intensively investigated over the past decades in all wavebands, i.e. from radio to very high energy (VHE) $\gamma$-rays. The radio observations of Coma Cluster have revealed an extended, non-thermal, radio halo associated with the cluster core  \citep{radiohalo1959Natur.183.1663L, radio1970MNRAS.151....1W, Radio1993ApJ...406..399G, radio2003A&A...397...53T, 2011MNRAS.412....2B, radiohalo2019SSRv..215...16V} and a diffuse radio relic with an extension of $\sim2$ Mpc at approximately the virial radius \citep{radiorelic1979ApJ...233..453J,relicradio1997A&A...321...55D,2011MNRAS.412....2B}.
Also, \citet{sizecloudradio2007ApJ...659..267K} have investigated a $\mathrm{\sim4\, Mpc}$ diameter of diffuse radio structure around Coma Cluster.

The recent X-ray observations showed a complex morphology towards the Coma cluster, such as a sub-group in the South-West part of the cluster connected to NGC 4839 and another sub-structure between the center of the Coma cluster and NGC 4839 \citep{2003A&A...400..811N}. Also, X-ray observations revealed various evidence of infall of galaxies, such as the group of NGC 4839 \citep[e.g.][]{2001A&A...365L..74N, 2020A&A...634A..30M}. More recently, the X-ray observation of the Coma cluster with the SRG/eROSITA telescope revealed a rich structure extending up to the virial radius of the cluster \citep{2021A&A...651A..41C}. They explained this by a recent merging of the  Coma cluster with the NGC 4839 group indicated by the detected faint X-ray bridge, which connects the galaxy group with the cluster.

In the HE $\gamma$-ray band, the Coma cluster was first studied by \textit{Fermi}-LAT using six years of data, showing two low significance structures within the viral radius \citep{comaFermiCollab2016ApJ...819..149A}. Then using more data of \textit{Fermi}-LAT,  \citet{fermicoma2018PhRvD..98f3006X} detected extended $\gamma$-ray emission from the direction of the Coma cluster in the 0.2-300 GeV band characterized with a soft spectral index of $\Gamma\sim2.7$. The detection of HE $\gamma$-ray excess towards the Coma cluster with \textit{Fermi}-LAT data was confirmed by \citet{2021A&A...648A..60A}, where the total $\gamma$-ray emission was explained by hadronic interactions in the intracluster gas.

In this paper, we present a detailed investigation of $\gamma$-ray emission in the direction of the Coma cluster using more than 12 years of \textit{Fermi}-LAT data. In Section 2, we describe the details of the \textit{Fermi}-LAT data analysis and, in section 3, we report the results. Finally, in Section 4, the discussion and conclusion of this work are presented.

\section{\textit{Fermi}-LAT data analysis}

\subsection{Data selection and background models}

We used 12.3 years (August 4, 2008 - June 1, 2021) of \textit{Fermi}-LAT Pass 8 ULTRACLEANVETO (`FRONT + BACK') class events \citep{atwood2013pass}. The selection of this event class was on account of the lowest level of CR contamination, that makes this class ideal for diffuse emission study, which is the main goal of this analysis. The data selection was within a region of interest (ROI) of $\mathrm{12^{\circ}}$ around the center of the Coma cluster at R.A., Dec = (194.95, 27.98) between 100 MeV and 1 TeV energies. We analysed the data with Fermipy (version 1.0.0; \citealt{2017ICRC...35..824W}) and \textit{Fermitools} software packages v2.0.0\footnote{\url{https://github.com/fermi-lat/Fermitools-conda/wiki}} following the standard procedure of the binned maximum-likelihood analysis technique with the $\mathrm{P8R3\_ULTRACLEANVETO\_V3}$ instrument response functions (IRFs). The data reduction was performed by using the recommended $\mathrm{(DATA\_QUAL>0)\&\&(LAT\_CONFIG==1)}$ data quality filters. And to reduce the contamination from the Earth's atmosphere, we employed a zenith angle cut of $\mathrm{90^{\circ}}$ to the events.
 In the spatial binning, we applied a pixel size of $\mathrm{0.1^{\circ}}$ and nine logarithmic energy bins per decade for the spectral binning. In the construction of the sky model, we used \textit{Fermi}-LAT sources from the second incremental version of the fourth catalog (4FGL-DR2; \citealt{2020arXiv200511208B}, \citealt{2020ApJS..247...33A}) lying up to $\mathrm{5^{\circ}}$ beyond the ROI. To account for the diffuse emission, we modeled the Galactic diffuse emission model $\mathrm{(gll{\_}iem{\_}v07.fits)}$ with isotropic component $\mathrm{(iso{\_}P8R3{\_}ULTRACLEANVETO{\_}V3{\_}v1.txt)}$ relevant to the ULTRACLEANVETO event class.

\begin{figure*}
 \centering
 \includegraphics[max width=\linewidth]{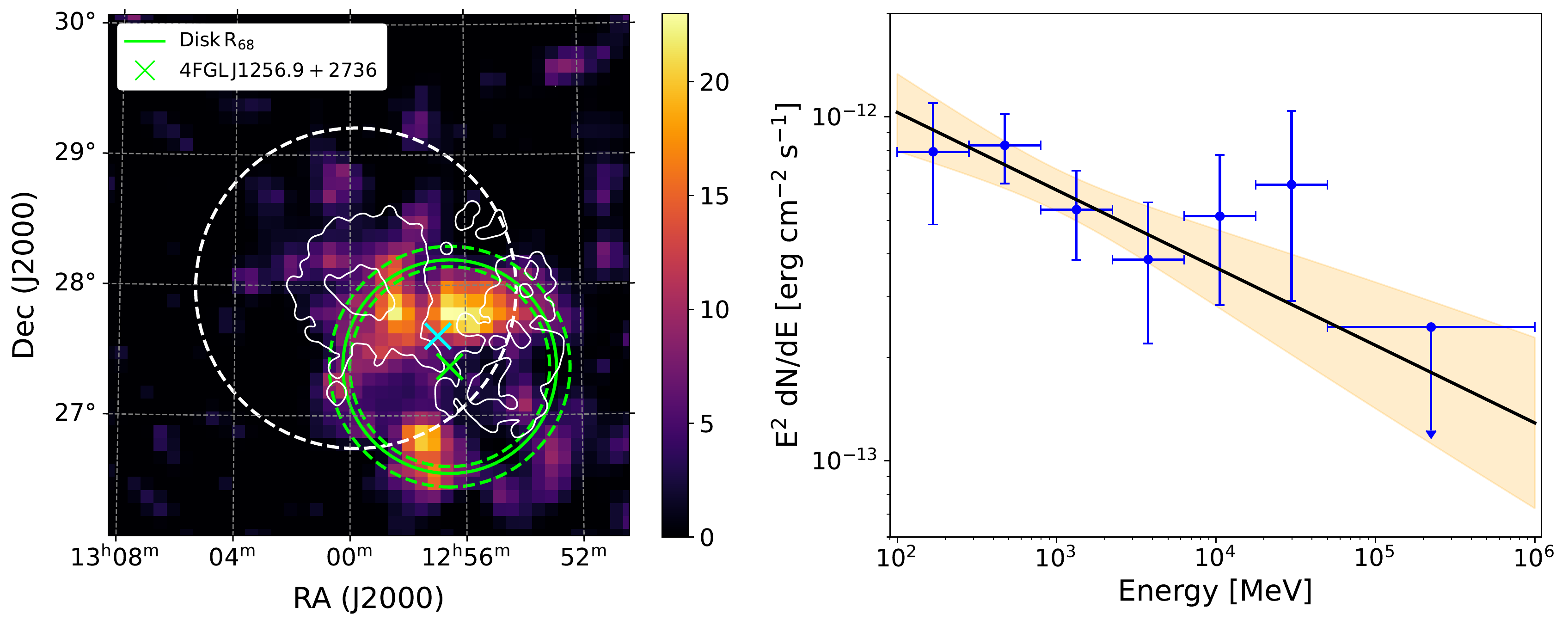}
\caption{Left: The TS map of the Coma cluster region at energies between 100 MeV and 1 TeV (left panel), where a green circle indicates the 68\% containment radius of a 2D homogeneous disk, while its corresponding $\mathrm{1\sigma}$ statistical errors are indicated by green dashed circles. Right: The SED of the region derived with the disk model, where the shaded orange area marks the $\mathrm{1\sigma}$ errors on the fitted spectral model. The ULs are computed at 95\% CL for the energy bins below $\mathrm{2\sigma}$.} 
\label{fig:fig2}
\end{figure*}
\raggedbottom

\subsection{Morphological and spectral analysis}

The detection of $\gamma$-ray emission from the direction of the Coma cluster has been already reported and studied by several authors using different spatial models \citep{fermicoma2018PhRvD..98f3006X,2021A&A...648A..60A,zargaryan_coma}. The excess $\gamma$-ray emission at (R.A., Dec) = (194.24,  27.61) degrees has been also added to the 4FGL-DR2 catalog as a point source named 4FGL J1256.9+2736 with power-law spectral index $\Gamma=2.73$. To study this excess, we analysed the data for 100 MeV-1 TeV, 100 MeV-1 GeV, and 1 GeV-1 TeV energy bands by removing 4FGL J1256.9+2736 from the source list as a baseline model. In the fitting, the spectral parameters of sources inside the ROI were left free, while for the sources outside the ROI, we used their 4FGL values, keeping them fixed during the fit. For all sources besides the diffuse background models, we applied the energy dispersion correction. After the first iteration of the fit, we removed sources with a test statistic (TS)\footnote{TS is defined as twice of the difference between the log-likelihoods of source plus background ($\mathrm{\mathcal{L}_1}$) and only background as a null hypothesis ($\mathrm{\mathcal{L}_0}$): $\mathrm{TS = 2log(\mathcal{L}_1/\mathcal{L}_0)}$.} lower than 1 and repeated the fitting until no source with $\mathrm{TS<1}$ was found. Then, we added the significant peaks above TS=16 beyond the  $\mathrm{2^{\circ}}$ from the ROI center to the model using \textit{find{\_}sources} tool implemented in Fermipy. This tool uses an iterative maximum-likelihood algorithm that creates a TS map of the ROI for the given model, identifies the significant peaks as new sources with the corresponding peak positions, and adds them to the model by repeatedly fitting. This process repeats until no peak is found above the given threshold. The TS maps of the resulting baseline models in the energy ranges of 100 MeV - 1 TeV, 100 MeV-1 GeV, and 1 GeV-1 TeV are shown in Figure \ref{fig:fig1} within $\mathrm{4^{\circ} \times 4^{\circ}}$ around the center of the Coma cluster. As shown in the TS map of the whole energy range (shown in the left panel of Figure \ref{fig:fig1}), the excess $\gamma$-ray emission has rather extended morphology shifted to the southwest from the virial radius of the Coma cluster. This structure remains similar in the TS map of the higher energy range (1 GeV - 1 TeV) with lower significance (see right panel of Figure \ref{fig:fig1}).

To explore the spectral and morphological characteristics of the excess $\gamma$-ray emission, first, we included 4FGL J1256.9+2736 in the model using the position in the 4FGL-DR2 catalog. Then, we tested the extension of the excess using the \textit{extension} tool in \textit{Fermipy}. This tool computes TS of the extension hypothesis defined as $\mathrm{TS_{ext}} = 2\times(\log\mathcal{L}_\mathrm{ext} - \log\mathcal{L}_\mathrm{PS})$, where $\mathrm{\log\mathcal{L}_{ext}}$ and $\mathrm{\log\mathcal{L}_{PS}}$ are likelihood values of the extended and point-like source models, respectively. As an input, we used the localised position of 4FGL J1256.9+2736 estimated with the \textit{localize} tool in \textit{Fermipy} after adding it into the baseline model. We checked the source extension for 2D Gaussian and 2D uniform-disk models, in both cases assuming a power-law spectral model. During the fitting, we left the position and the spectral parameters of the testing source and the normalization of all sources within $\mathrm{5^\circ}$ from the center of ROI to vary. For the 2D uniform disk model we obtained $\mathrm{TS_{ext}\sim29.3}$ with 68\% containment radius of $\mathrm{R_{68}=0.82^{+0.10}_{-0.05}}$, which is compatible with the significance map between 100 MeV and 1 GeV, shown in the middle panel of \ref{fig:fig1}. With the 2D symmetric Gaussian model this test showed no significant extension ($\mathrm{TS_{ext}\sim9.8}$). The results of the extension analysis are summarized in Table \ref{tab:table1} and the detected extension with a uniform disk model is shown in the left upper panel of Figure \ref{fig:fig2}. 

Also, we take into account the possibility that the observed extended morphology can be a superposition of several point-like sources such as AGNs in the cluster or at least have a contribution from them. To check this, we applied the \textit{find{\_}sources} source-finding algorithm on the excess emission, which identified three point-like sources at localised positions of $\mathrm{(R.A., Dec) = (194.63, 27.83)}$ degrees, $\mathrm{(R.A., Dec) = (193.86, 27.82)}$ degrees, and $\mathrm{(R.A., Dec) = (194.37, 26.82)}$ degrees with detection significances above $\mathrm{\sim 4\sigma}$. Hereafter, these point-like sources will be referred to as $\mathrm{p_1}$, $\mathrm{p_2}$, and $\mathrm{p_3}$, respectively. The corresponding 68\% containment regions are shown in the left panel of Figure \ref{fig:fig1}. Then, we considered two multi-source models, that are the combination of the identified point-like sources as $\mathrm{p_1+p_2+p_3}$ and $\mathrm{disk+p_1+p_2+p_3}$, which also includes the 2D uniform disk model obtained from the extension analysis. In the considered models, all components were described by a power-law spectral model.

\begin{figure*}
 \centering
\includegraphics[width=\textwidth]{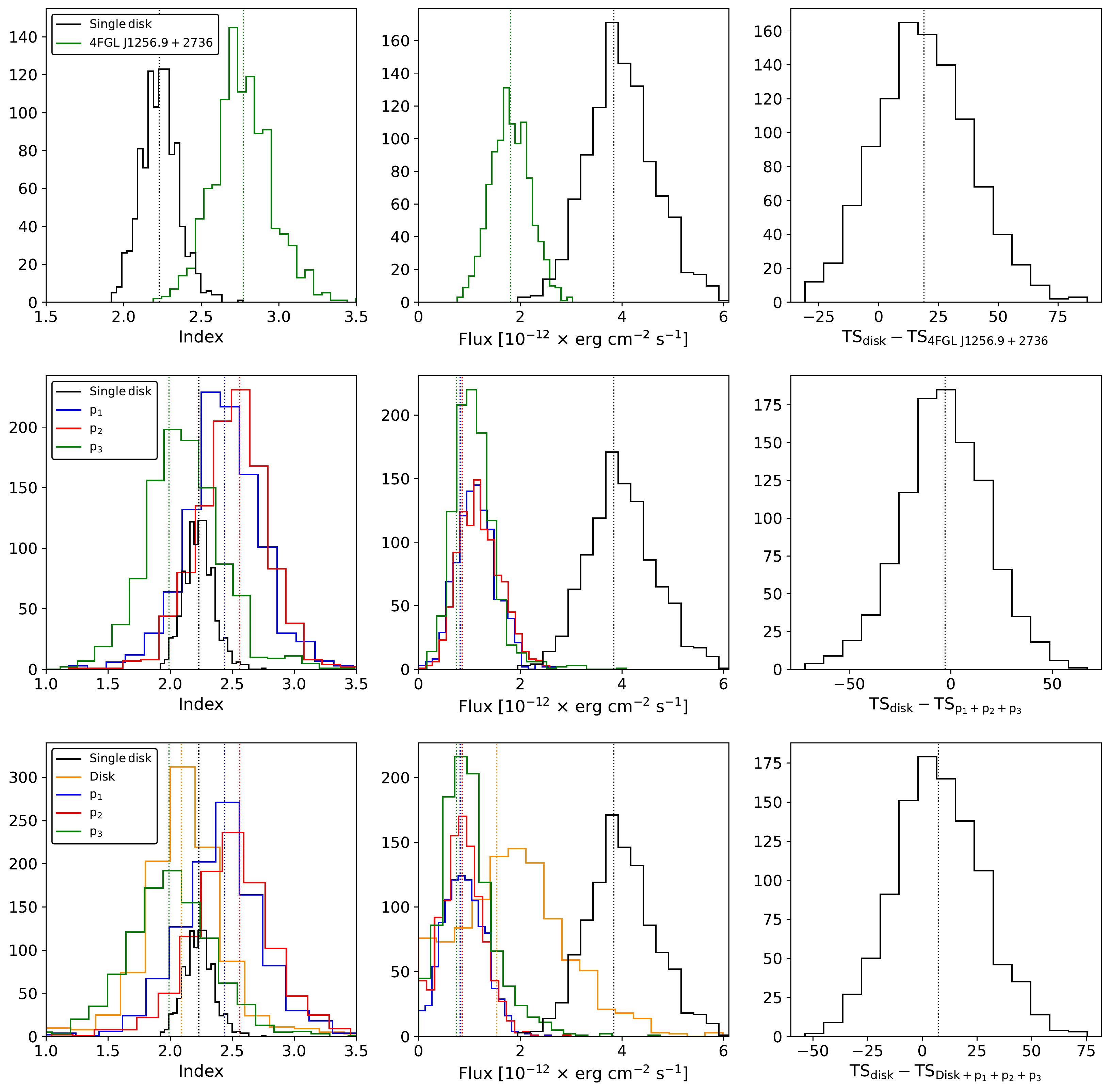}
\caption{Distribution of spectral indices, energy fluxes, and TS values of 4FGL J1256.9+2736 (upper panels), $\mathrm{p_1+p_2+p_3}$ (middle panels), and $\mathrm{disk+p_1+p_2+p_3}$ (lower panels) models in comparison with the disk model derived from the fitting of the 1000 simulated data. Dashed lines in the index and flux distributions correspond to the initial values used as inputs to the simulations.  The black dashed lines in the right panels indicate the median values of the distributions.}
\label{fig:fig3}
\end{figure*}

\section{Results}

A spectral energy distribution (SED) of the Coma cluster region corresponding to the disk model extends up to 50 GeV (see Figure \ref{fig:fig2}). It is well fitted with a power-law index $\mathrm{\Gamma=2.23\pm0.11}$ and a total flux of $\mathrm{(5.11\pm1.15)\times10^{-9}\,ph\,cm^{-2}\,s^{-1}}$, which agrees well with the results reported in \citet{comaFermiCollab2016ApJ...819..149A}. In particular, for the cored profile and homogeneous disk models, they estimate upper limits (ULs) of about $\mathrm{(5.2-5.8)\times10^{-9}\,ph\,cm^{-2}\,s^{-1}}$ above 100 MeV with a power-law spectral index $\mathrm{\Gamma\sim2.3}$.

To compare our results with the recent studies of the Coma cluster with Fermi-LAT, additionally, we also applied the disk model in the energy range between 200 MeV and 300 GeV. It resulted in spectral index of $\mathrm{\Gamma=2.24\pm0.12}$ and photon flux $\mathrm{(2.41\pm0.45)\times10^{-9}\,ph\,cm^{-2}\,s^{-1}}$. The flux is compatible to the total flux of the extended models found in \citet{fermicoma2018PhRvD..98f3006X} for the same energy range which is around $\mathrm{(2.3-3.1)\times10^{-9}\,ph\,cm^{-2}\,s^{-1}}$. The flux of the best-fit model reported by \citet{2021A&A...648A..60A} when 4FGL J1256.9+2736 is not included in the model is about $\mathrm{1.5\times10^{-9}\,ph\,cm^{-2}\,s^{-1}}$, which is slightly lower than our total flux for the same energy range. Also, the spectral index we obtained for the disk model is significantly harder than what was found in \citet{2021A&A...648A..60A} by using an underlying physical model in the fit, while our model is data-driven.

The best-fit parameters of the models, described in the previous section, are given in Table \ref{tab:table2} with the corresponding log-likelihood ($log\mathcal{L}$), and the Akaike information criterion (AIC, \citep{ 1974ITAC...19..716A}) values. The AIC is defined as $\mathrm{AIC=2(\Delta d.o.f.-\Delta \log\mathcal{L})}$, where the difference between the number of degrees of freedom (DOF) ($\mathrm{\Delta d.o.f.}$) and the log-likelihood ($\mathrm{\Delta \log\mathcal{L}}$) values are computed relative to the homogeneous disk model. In this formulation, the models with the smallest AIC value are considered the favored model, however, we should note that the AIC method does not allow estimating the significance of the differences between tested models relative to each other.

With respect to $\mathrm{\Delta AIC}$ estimates listed in table \ref{tab:table2}, we find that $\mathrm{disk+p_1+p_2+p_3}$ and $\mathrm{p_1+p_2+p_3}$ have preferential values over the single disk model with -7.2 and -4.7, respectively. The results indicate that the diffuse emission can have a contribution from three candidates of point-like sources or the extended emission is itself a combination of those three candidates. Additionally, we replaced the disk model with the 4FGLJ1256.9+2736 source across the entire region using the position in the 4FGL-DR2 catalog. For this case, the $\mathrm{\Delta AIC}$ relative to the disk model is 21.2, which indicates that 4FGLJ1256.9+2736 can not independently describe the region and that an additional component is needed.

To reduce the possibility of being misled by potential fluctuations in the models, we conducted spectral analysis on 1000 simulated datasets between 100 MeV and 1 TeV using the best-fit parameters of the model components as initial values (see Appendix A for a more detailed description). On Figure \ref{fig:fig3}, we show the distribution of the spectral indices, energy fluxes, and TS values for 4FGL J1256.9+2736 (upper panel), $\mathrm{p_1+p_2+p_3}$ (middle panel), and $\mathrm{disk+p_1+p_2+p_3}$ (lower panel) models in comparison with the disk model. While the models are not nested, we expect more accurate models to result in larger TS in the region of the Coma cluster. We can see this based on the comparison of the disk and 4FGL J1256.9+2736 models displayed in Figure \ref{fig:fig3}, meaning that the disk model describes the region much better than 4FGL J1256.9+2736 ($\mathrm{TS_{disk}-TS_{4FGL\,J1256.9+2736}\sim19}$).  
A similar comparison for the $\mathrm{p_1+p_2+p_3}$ model (middle right panel) result in $\mathrm{TS_{disk}-TS_{p_1+p_2+p_3}\sim-3}$, meaning that three point-like sources explain the region slightly more accurately than a single disk component. By comparison of $\mathrm{disk+p_1+p_2+p_3}$ with the disk model, we get $\mathrm{TS_{disk}-TS_{disk+p_1+p_2+p_3}\sim7}$, indicating that the inclusion of sources like points results in an underestimation of the region. We notice, however, that the difference is not significant in both cases, so no conclusion can be drawn from this.

We also evaluated the systematic uncertainties in the spectral parameters of the disk model due to the choice of Galactic interstellar emission model (IEM), the event class, and the low energy threshold. The results are summarised in Table \ref{tab:table3}. The systematic uncertainties on the spectral index due to the selection of IEM is $\mathrm{<5\%}$ and $\mathrm{<18\%}$ for the flux. The systematic errors coming from the selection of event classes are estimated to be $\mathrm{<15\%}$ on the energy flux and $\mathrm{<3\%}$ on the spectra index. The change in the low energy threshold from 100 MeV to 500 MeV results in a spectral index uncertainty of $\mathrm{<5\%}$, while the impact on the flux is $\mathrm{<86\%}$, suggesting that a sufficient fraction of the data is produced below 500 MeV.

\begin{table*}
\caption{Summary of the main systematic uncertainties estimated for the disk model.}
\centering
\label{tab:table3}
\begin{threeparttable}
\begin{tabular}{lcccccc}
    \hline
    Type & Details & Impact on flux & Impact on energy flux& Impact on index & Variation in TS & Variation in $\mathrm{N_{pred}}$ \\
    \hline
    Alt. IEMs & Alt. IEMs models & <18\% & <28\% & <5\% & 44-69 & 730-889\\
    Alt. event classes & Alt. event class & <15\% & <7\%& <3\% & 63-65 & 1252-1310\\
    Low energy threshold & 100-500 MeV & <86\% & <32\%& <5\%  & 47-52 & 268-863\\
    \hline
\end{tabular}
\end{threeparttable}
\end{table*}

\section{Discussion and conclusion}

We have presented the detection of diffuse $\gamma$-ray emission in the vicinity of the Coma cluster at the level of $\mathrm{\sim7.2\sigma}$ and $\mathrm{\sim5.4\sigma}$ extension significance assuming a 2D homogeneous disk model. The localised position of the extension center ($\mathrm{RA=194.14\pm0.14, Dec=27.38\pm0.13}$ degrees) has an offset of $\mathrm{0.94^{\circ}}$ from the center of the viral region of the Coma cluster to the south-
west, while the radius $\mathrm{r=1.0^{+0.12}_{-0.06}}$ degrees ($\mathrm{R_{68}=0.82r}$) is compatible with the virial radius  of the cluster ($\mathrm{\Theta_{200}=1.23^{\circ}}$). The corresponding spectrum of the excess extends up to $\sim50$ GeV exhibiting a hard power-law spectrum with $\mathrm{\Gamma=2.23\pm0.11}$ and energy flux of $\mathrm{(3.84\pm0.67)\times10^{-12}\,erg\,cm^{-2}\,s^{-1}}$.  By comparing the disk model with point source 4FGL J1256.9+2736, which in the 4FGL-DR2 catalog is associated with NGC 4839, we showed, that the observed $\gamma$-ray emission can not be explained by a point source (see the previous section). As we have already mentioned, the diffuse emission could be a product of several point-like or extended sources, or at least there can be a contribution from them. Unfortunately because of the poor angular resolution of Fermi-LAT below 1 GeV as well as low statistics, we could not reject or confirm such a possibility statistically (see the previous section for more details). 

The main candidates of the $\gamma$-ray emission are radiogalaxies and star-forming galaxies. The former was estimated by \citet{comaFermiCollab2016ApJ...819..149A} for the radiogalaxies NGC 4869 and NGC 4874. Both of these sources are located within the $\gamma$-ray diffuse emission and according to \citet{1985MNRAS.215..437C}, they are the radio brightest radiogalaxies in the cluster. The combined $\gamma$-ray luminosity was estimated to be $\mathrm{L_{\gamma}\sim8\times10^{40}\,erg\,s^{-1}}$, assuming that the $\gamma$-ray emission is produced via IC scattering of CMB photons by radio-producing electrons (IC/CMB model). This is only the $\mathrm{\sim2\%}$ of the observed $\gamma$-ray luminosity assuming the disk model, which is $L_{\gamma}=(4.6\pm0.8)\times10^{42}\, \mathrm{erg \,s^{-1}}$. Other radiogalaxies in the cluster and the background, such as NGC 4839, IC 3960, IC 3959, and NGC 4850, have weaker radio fluxes; therefore, one can conclude that they can not be responsible for the whole observed diffuse $\gamma$-ray emission. The expected $\gamma$-ray emission from star-forming galaxies in 0.1-100 GeV range was estimated by \citet{2012ApJ...755..117S} to be $L_{\gamma}\mathrm{\sim3\times10^{40}}-\mathrm{3\times10^{42} \,erg\, s^{-1}}$ range. Taking into account the fact that the detected $\gamma$-ray emission has an extension of $\mathrm{\sim5}$ Mpc from the North
to the South, it is less likely that star-forming galaxies can provide the observed morphology, even if the higher limit of the estimated $\gamma$-ray luminosity is on the order of the observed one. 

Very recently, \citet{2021A&A...651A..41C} reported the first results of the X-ray studies of SRG/eROSITA X-ray studies of the Coma cluster. They revealed a complex morphology with two structures around the core, one of which is to the East with $\mathrm{\sim0.5}$ Mpc in size and a bow-like sharp X-ray feature to the West with $\mathrm{\sim3}$ Mpc size in North to South direction. Also, they found a faint X-ray bridge connecting the NGC 4839 group with the cluster. Based on this, they assume that the two extended X-ray features can be the results of a recent merger of the cluster with the NGC 4839 group as discussed in \citet{2021MNRAS.501.1038Z}. The total X-ray extension of the Coma cluster including the NGC 4839 group was found to be about $\mathrm{\sim3}$ Mpc from the North to the South. In comparison, the $\gamma$-ray extension that we found is around $\mathrm{\sim4}$ Mpc in the North to the South direction. However, to claim a correlation with the X-ray morphology found by \citet{2021A&A...651A..41C} or resolve the observed morphology more clearly, we need considerably more data.

\section*{Acknowledgements}
DZ acknowledges funding from an Irish Research Council (IRC) Starting Laureate Award (IRCLA\textbackslash 2017\textbackslash 83).
JM acknowledges support from a Royal Society-Science Foundation Ireland University Research Fellowship (14/RS-URF/3219, 20/RS-URF-R/3712) and an Irish Research Council (IRC) Starting Laureate Award (IRCLA\textbackslash 2017\textbackslash 83). SC acknowledges funding from the Polish Science Centre grant DEC-2017/27/B/ST9/02272.  
\section*{Data availability}

All of the data underlying this article are publicly available and can be downloaded from the \textit{Fermi}-LAT Data Server.

\bibliographystyle{mnras}
\bibliography{bibliography}

\begin{thebibliography}{}
\makeatletter
\relax
\def\mn@urlcharsother{\let\do\@makeother \do\$\do\&\do\#\do\^\do\_\do\%\do\~}
\def\mn@doi{\begingroup\mn@urlcharsother \@ifnextchar [ {\mn@doi@}
  {\mn@doi@[]}}
\def\mn@doi@[#1]#2{\def\@tempa{#1}\ifx\@tempa\@empty \href
  {http://dx.doi.org/#2} {doi:#2}\else \href {http://dx.doi.org/#2} {#1}\fi
  \endgroup}
\def\mn@eprint#1#2{\mn@eprint@#1:#2::\@nil}
\def\mn@eprint@arXiv#1{\href {http://arxiv.org/abs/#1} {{\tt arXiv:#1}}}
\def\mn@eprint@dblp#1{\href {http://dblp.uni-trier.de/rec/bibtex/#1.xml}
  {dblp:#1}}
\def\mn@eprint@#1:#2:#3:#4\@nil{\def\@tempa {#1}\def\@tempb {#2}\def\@tempc
  {#3}\ifx \@tempc \@empty \let \@tempc \@tempb \let \@tempb \@tempa \fi \ifx
  \@tempb \@empty \def\@tempb {arXiv}\fi \@ifundefined
  {mn@eprint@\@tempb}{\@tempb:\@tempc}{\expandafter \expandafter \csname
  mn@eprint@\@tempb\endcsname \expandafter{\@tempc}}}

\bibitem[\protect\citeauthoryear{{Abdollahi} et~al.,}{{Abdollahi}
  et~al.}{2020}]{2020ApJS..247...33A}
{Abdollahi} S.,  et~al., 2020, \mn@doi [\apjs] {10.3847/1538-4365/ab6bcb},
  \href {https://ui.adsabs.harvard.edu/abs/2020ApJS..247...33A} {247, 33}

\bibitem[\protect\citeauthoryear{{Acero} et~al.,}{{Acero}
  et~al.}{2016}]{2016ApJS..224....8A}
{Acero} F.,  et~al., 2016, \mn@doi [\apjs] {10.3847/0067-0049/224/1/8}, \href
  {https://ui.adsabs.harvard.edu/abs/2016ApJS..224....8A} {224, 8}

\bibitem[\protect\citeauthoryear{{Ackermann} et~al.,}{{Ackermann}
  et~al.}{2016}]{comaFermiCollab2016ApJ...819..149A}
{Ackermann} M.,  et~al., 2016, \mn@doi [\apj] {10.3847/0004-637X/819/2/149},
  \href {https://ui.adsabs.harvard.edu/abs/2016ApJ...819..149A} {819, 149}

\bibitem[\protect\citeauthoryear{{Adam}, {Goksu}, {Brown}, {Rudnick}  \&
  {Ferrari}}{{Adam} et~al.}{2021}]{2021A&A...648A..60A}
{Adam} R.,  {Goksu} H.,  {Brown} S.,  {Rudnick} L.,   {Ferrari} C.,  2021,
  \mn@doi [\aap] {10.1051/0004-6361/202039660}, \href
  {https://ui.adsabs.harvard.edu/abs/2021A&A...648A..60A} {648, A60}

\bibitem[\protect\citeauthoryear{{Aharonian}}{{Aharonian}}{2002}]{2002MNRAS.332..215A}
{Aharonian} F.~A.,  2002, \mn@doi [\mnras] {10.1046/j.1365-8711.2002.05292.x},
  \href {https://ui.adsabs.harvard.edu/abs/2002MNRAS.332..215A} {332, 215}

\bibitem[\protect\citeauthoryear{{Akaike}}{{Akaike}}{1974}]{1974ITAC...19..716A}
{Akaike} H.,  1974, IEEE Transactions on Automatic Control, \href
  {https://ui.adsabs.harvard.edu/abs/1974ITAC...19..716A} {19, 716}

\bibitem[\protect\citeauthoryear{{Atoyan} \& {V{\"o}lk}}{{Atoyan} \&
  {V{\"o}lk}}{2000}]{2000ApJ...535...45A}
{Atoyan} A.~M.,  {V{\"o}lk} H.~J.,  2000, \mn@doi [\apj] {10.1086/308828},
  \href {https://ui.adsabs.harvard.edu/abs/2000ApJ...535...45A} {535, 45}

\bibitem[\protect\citeauthoryear{Atwood et~al.,}{Atwood
  et~al.}{2013}]{atwood2013pass}
Atwood W.,  et~al., 2013, Pass 8: Toward the Full Realization of the Fermi-LAT
  Scientific Potential (\mn@eprint {arXiv} {1303.3514})

\bibitem[\protect\citeauthoryear{{Ballet}, {Burnett}, {Digel}  \&
  {Lott}}{{Ballet} et~al.}{2020}]{2020arXiv200511208B}
{Ballet} J.,  {Burnett} T.~H.,  {Digel} S.~W.,   {Lott} B.,  2020, arXiv
  e-prints, \href {https://ui.adsabs.harvard.edu/abs/2020arXiv200511208B} {p.
  arXiv:2005.11208}

\bibitem[\protect\citeauthoryear{{Berezinsky}, {Blasi}  \&
  {Ptuskin}}{{Berezinsky} et~al.}{1997}]{1997ApJ...487..529B}
{Berezinsky} V.~S.,  {Blasi} P.,   {Ptuskin} V.~S.,  1997, \mn@doi [\apj]
  {10.1086/304622}, \href
  {https://ui.adsabs.harvard.edu/abs/1997ApJ...487..529B} {487, 529}

\bibitem[\protect\citeauthoryear{{Blasi}, {Gabici}  \& {Brunetti}}{{Blasi}
  et~al.}{2007}]{2007IJMPA..22..681B}
{Blasi} P.,  {Gabici} S.,   {Brunetti} G.,  2007, \mn@doi [International
  Journal of Modern Physics A] {10.1142/S0217751X0703529X}, \href
  {https://ui.adsabs.harvard.edu/abs/2007IJMPA..22..681B} {22, 681}

\bibitem[\protect\citeauthoryear{{Bonafede}, {Feretti}, {Murgia}, {Govoni},
  {Giovannini}, {Dallacasa}, {Dolag}  \& {Taylor}}{{Bonafede}
  et~al.}{2010}]{2010A&A...513A..30B}
{Bonafede} A.,  {Feretti} L.,  {Murgia} M.,  {Govoni} F.,  {Giovannini} G.,
  {Dallacasa} D.,  {Dolag} K.,   {Taylor} G.~B.,  2010, \mn@doi [\aap]
  {10.1051/0004-6361/200913696}, \href
  {https://ui.adsabs.harvard.edu/abs/2010A&A...513A..30B} {513, A30}

\bibitem[\protect\citeauthoryear{{Brown} \& {Rudnick}}{{Brown} \&
  {Rudnick}}{2011}]{2011MNRAS.412....2B}
{Brown} S.,  {Rudnick} L.,  2011, \mn@doi [\mnras]
  {10.1111/j.1365-2966.2010.17738.x}, \href
  {https://ui.adsabs.harvard.edu/abs/2011MNRAS.412....2B} {412, 2}

\bibitem[\protect\citeauthoryear{{Brunetti} \& {Blasi}}{{Brunetti} \&
  {Blasi}}{2005}]{2005MNRAS.363.1173B}
{Brunetti} G.,  {Blasi} P.,  2005, \mn@doi [\mnras]
  {10.1111/j.1365-2966.2005.09511.x}, \href
  {https://ui.adsabs.harvard.edu/abs/2005MNRAS.363.1173B} {363, 1173}

\bibitem[\protect\citeauthoryear{{Carilli} \& {Taylor}}{{Carilli} \&
  {Taylor}}{2002}]{2002ARA&A..40..319C}
{Carilli} C.~L.,  {Taylor} G.~B.,  2002, \mn@doi [\araa]
  {10.1146/annurev.astro.40.060401.093852}, \href
  {https://ui.adsabs.harvard.edu/abs/2002ARA&A..40..319C} {40, 319}

\bibitem[\protect\citeauthoryear{{Churazov}, {Khabibullin}, {Lyskova},
  {Sunyaev}  \& {Bykov}}{{Churazov} et~al.}{2021}]{2021A&A...651A..41C}
{Churazov} E.,  {Khabibullin} I.,  {Lyskova} N.,  {Sunyaev} R.,   {Bykov}
  A.~M.,  2021, \mn@doi [\aap] {10.1051/0004-6361/202040197}, \href
  {https://ui.adsabs.harvard.edu/abs/2021A&A...651A..41C} {651, A41}

\bibitem[\protect\citeauthoryear{{Colafrancesco} \& {Blasi}}{{Colafrancesco} \&
  {Blasi}}{1998}]{1998APh.....9..227C}
{Colafrancesco} S.,  {Blasi} P.,  1998, \mn@doi [Astroparticle Physics]
  {10.1016/S0927-6505(98)00018-8}, \href
  {https://ui.adsabs.harvard.edu/abs/1998APh.....9..227C} {9, 227}

\bibitem[\protect\citeauthoryear{{Cordey}}{{Cordey}}{1985}]{1985MNRAS.215..437C}
{Cordey} R.~A.,  1985, \mn@doi [\mnras] {10.1093/mnras/215.3.437}, \href
  {https://ui.adsabs.harvard.edu/abs/1985MNRAS.215..437C} {215, 437}

\bibitem[\protect\citeauthoryear{{Deiss}, {Reich}, {Lesch}  \&
  {Wielebinski}}{{Deiss} et~al.}{1997}]{relicradio1997A&A...321...55D}
{Deiss} B.~M.,  {Reich} W.,  {Lesch} H.,   {Wielebinski} R.,  1997, \aap, \href
  {https://ui.adsabs.harvard.edu/abs/1997A&A...321...55D} {321, 55}

\bibitem[\protect\citeauthoryear{{Dennison}}{{Dennison}}{1980}]{1980ApJ...239L..93D}
{Dennison} B.,  1980, \mn@doi [\apjl] {10.1086/183300}, \href
  {https://ui.adsabs.harvard.edu/abs/1980ApJ...239L..93D} {239, L93}

\bibitem[\protect\citeauthoryear{{Donnert}, {Vazza}, {Br{\"u}ggen}  \&
  {ZuHone}}{{Donnert} et~al.}{2018}]{2018SSRv..214..122D}
{Donnert} J.,  {Vazza} F.,  {Br{\"u}ggen} M.,   {ZuHone} J.,  2018, \mn@doi
  [\ssr] {10.1007/s11214-018-0556-8}, \href
  {https://ui.adsabs.harvard.edu/abs/2018SSRv..214..122D} {214, 122}

\bibitem[\protect\citeauthoryear{{Ensslin}, {Biermann}, {Kronberg}  \&
  {Wu}}{{Ensslin} et~al.}{1997}]{1997ApJ...477..560E}
{Ensslin} T.~A.,  {Biermann} P.~L.,  {Kronberg} P.~P.,   {Wu} X.-P.,  1997,
  \mn@doi [\apj] {10.1086/303722}, \href
  {https://ui.adsabs.harvard.edu/abs/1997ApJ...477..560E} {477, 560}

\bibitem[\protect\citeauthoryear{{Gabici} \& {Blasi}}{{Gabici} \&
  {Blasi}}{2003a}]{2003APh....19..679G}
{Gabici} S.,  {Blasi} P.,  2003a, \mn@doi [Astroparticle Physics]
  {10.1016/S0927-6505(03)00106-3}, \href
  {https://ui.adsabs.harvard.edu/abs/2003APh....19..679G} {19, 679}

\bibitem[\protect\citeauthoryear{{Gabici} \& {Blasi}}{{Gabici} \&
  {Blasi}}{2003b}]{2003ApJ...583..695G}
{Gabici} S.,  {Blasi} P.,  2003b, \mn@doi [\apj] {10.1086/345429}, \href
  {https://ui.adsabs.harvard.edu/abs/2003ApJ...583..695G} {583, 695}

\bibitem[\protect\citeauthoryear{{Gabici} \& {Blasi}}{{Gabici} \&
  {Blasi}}{2004}]{2004APh....20..579G}
{Gabici} S.,  {Blasi} P.,  2004, \mn@doi [Astroparticle Physics]
  {10.1016/j.astropartphys.2003.09.002}, \href
  {https://ui.adsabs.harvard.edu/abs/2004APh....20..579G} {20, 579}

\bibitem[\protect\citeauthoryear{{Giovannini} \& {Feretti}}{{Giovannini} \&
  {Feretti}}{2000}]{2000NewA....5..335G}
{Giovannini} G.,  {Feretti} L.,  2000, \mn@doi [\na]
  {10.1016/S1384-1076(00)00034-8}, \href
  {https://ui.adsabs.harvard.edu/abs/2000NewA....5..335G} {5, 335}

\bibitem[\protect\citeauthoryear{{Giovannini}, {Feretti}, {Venturi}, {Kim}  \&
  {Kronberg}}{{Giovannini} et~al.}{1993}]{Radio1993ApJ...406..399G}
{Giovannini} G.,  {Feretti} L.,  {Venturi} T.,  {Kim} K.~T.,   {Kronberg}
  P.~P.,  1993, \mn@doi [\apj] {10.1086/172451}, \href
  {https://ui.adsabs.harvard.edu/abs/1993ApJ...406..399G} {406, 399}

\bibitem[\protect\citeauthoryear{{Govoni} \& {Feretti}}{{Govoni} \&
  {Feretti}}{2004}]{2004IJMPD..13.1549G}
{Govoni} F.,  {Feretti} L.,  2004, \mn@doi [International Journal of Modern
  Physics D] {10.1142/S0218271804005080}, \href
  {https://ui.adsabs.harvard.edu/abs/2004IJMPD..13.1549G} {13, 1549}

\bibitem[\protect\citeauthoryear{{Hillas}}{{Hillas}}{1984}]{1984ARA&A..22..425H}
{Hillas} A.~M.,  1984, \mn@doi [\araa] {10.1146/annurev.aa.22.090184.002233},
  \href {https://ui.adsabs.harvard.edu/abs/1984ARA&A..22..425H} {22, 425}

\bibitem[\protect\citeauthoryear{{Hinton}, {Domainko}  \& {Pope}}{{Hinton}
  et~al.}{2007}]{2007MNRAS.382..466H}
{Hinton} J.~A.,  {Domainko} W.,   {Pope} E.~C.~D.,  2007, \mn@doi [\mnras]
  {10.1111/j.1365-2966.2007.12395.x}, \href
  {https://ui.adsabs.harvard.edu/abs/2007MNRAS.382..466H} {382, 466}

\bibitem[\protect\citeauthoryear{{Inoue}, {Aharonian}  \& {Sugiyama}}{{Inoue}
  et~al.}{2005}]{2005ApJ...628L...9I}
{Inoue} S.,  {Aharonian} F.~A.,   {Sugiyama} N.,  2005, \mn@doi [\apjl]
  {10.1086/432602}, \href
  {https://ui.adsabs.harvard.edu/abs/2005ApJ...628L...9I} {628, L9}

\bibitem[\protect\citeauthoryear{{Jaffe} \& {Rudnick}}{{Jaffe} \&
  {Rudnick}}{1979}]{radiorelic1979ApJ...233..453J}
{Jaffe} W.~J.,  {Rudnick} L.,  1979, \mn@doi [\apj] {10.1086/157406}, \href
  {https://ui.adsabs.harvard.edu/abs/1979ApJ...233..453J} {233, 453}

\bibitem[\protect\citeauthoryear{{Kelner} \& {Aharonian}}{{Kelner} \&
  {Aharonian}}{2008}]{2008PhRvD..78c4013K}
{Kelner} S.~R.,  {Aharonian} F.~A.,  2008, \mn@doi [\prd]
  {10.1103/PhysRevD.78.034013}, \href
  {https://ui.adsabs.harvard.edu/abs/2008PhRvD..78c4013K} {78, 034013}

\bibitem[\protect\citeauthoryear{{Kronberg}, {Kothes}, {Salter}  \&
  {Perillat}}{{Kronberg} et~al.}{2007}]{sizecloudradio2007ApJ...659..267K}
{Kronberg} P.~P.,  {Kothes} R.,  {Salter} C.~J.,   {Perillat} P.,  2007,
  \mn@doi [\apj] {10.1086/511512}, \href
  {https://ui.adsabs.harvard.edu/abs/2007ApJ...659..267K} {659, 267}

\bibitem[\protect\citeauthoryear{{Large}, {Mathewson}  \& {Haslam}}{{Large}
  et~al.}{1959}]{radiohalo1959Natur.183.1663L}
{Large} M.~I.,  {Mathewson} D.~S.,   {Haslam} C.~G.~T.,  1959, \mn@doi [\nat]
  {10.1038/1831663a0}, \href
  {https://ui.adsabs.harvard.edu/abs/1959Natur.183.1663L} {183, 1663}

\bibitem[\protect\citeauthoryear{{Malavasi}, {Aghanim}, {Tanimura}, {Bonjean}
  \& {Douspis}}{{Malavasi} et~al.}{2020}]{2020A&A...634A..30M}
{Malavasi} N.,  {Aghanim} N.,  {Tanimura} H.,  {Bonjean} V.,   {Douspis} M.,
  2020, \mn@doi [\aap] {10.1051/0004-6361/201936629}, \href
  {https://ui.adsabs.harvard.edu/abs/2020A&A...634A..30M} {634, A30}

\bibitem[\protect\citeauthoryear{{Neumann} et~al.,}{{Neumann}
  et~al.}{2001}]{2001A&A...365L..74N}
{Neumann} D.~M.,  et~al., 2001, \mn@doi [\aap] {10.1051/0004-6361:20000182},
  \href {https://ui.adsabs.harvard.edu/abs/2001A&A...365L..74N} {365, L74}

\bibitem[\protect\citeauthoryear{{Neumann}, {Lumb}, {Pratt}  \&
  {Briel}}{{Neumann} et~al.}{2003}]{2003A&A...400..811N}
{Neumann} D.~M.,  {Lumb} D.~H.,  {Pratt} G.~W.,   {Briel} U.~G.,  2003, \mn@doi
  [\aap] {10.1051/0004-6361:20021911}, \href
  {https://ui.adsabs.harvard.edu/abs/2003A&A...400..811N} {400, 811}

\bibitem[\protect\citeauthoryear{{Planck Collaboration} et~al.,}{{Planck
  Collaboration} et~al.}{2013}]{2013A&A...554A.140P}
{Planck Collaboration} et~al., 2013, \mn@doi [\aap]
  {10.1051/0004-6361/201220247}, \href
  {https://ui.adsabs.harvard.edu/abs/2013A&A...554A.140P} {554, A140}

\bibitem[\protect\citeauthoryear{{Ryu}, {Kang}, {Hallman}  \& {Jones}}{{Ryu}
  et~al.}{2003}]{2003ApJ...593..599R}
{Ryu} D.,  {Kang} H.,  {Hallman} E.,   {Jones} T.~W.,  2003, \mn@doi [\apj]
  {10.1086/376723}, \href
  {https://ui.adsabs.harvard.edu/abs/2003ApJ...593..599R} {593, 599}

\bibitem[\protect\citeauthoryear{{Storm}, {Jeltema}  \& {Profumo}}{{Storm}
  et~al.}{2012}]{2012ApJ...755..117S}
{Storm} E.~M.,  {Jeltema} T.~E.,   {Profumo} S.,  2012, \mn@doi [\apj]
  {10.1088/0004-637X/755/2/117}, \href
  {https://ui.adsabs.harvard.edu/abs/2012ApJ...755..117S} {755, 117}

\bibitem[\protect\citeauthoryear{{Struble} \& {Rood}}{{Struble} \&
  {Rood}}{1991}]{redshift1991ApJS...77..363S}
{Struble} M.~F.,  {Rood} H.~J.,  1991, \mn@doi [\apjs] {10.1086/191608}, \href
  {https://ui.adsabs.harvard.edu/abs/1991ApJS...77..363S} {77, 363}

\bibitem[\protect\citeauthoryear{{Thierbach}, {Klein}  \&
  {Wielebinski}}{{Thierbach} et~al.}{2003}]{radio2003A&A...397...53T}
{Thierbach} M.,  {Klein} U.,   {Wielebinski} R.,  2003, \mn@doi [\aap]
  {10.1051/0004-6361:20021474}, \href
  {https://ui.adsabs.harvard.edu/abs/2003A&A...397...53T} {397, 53}

\bibitem[\protect\citeauthoryear{{Timokhin}, {Aharonian}  \&
  {Neronov}}{{Timokhin} et~al.}{2004}]{2004A&A...417..391T}
{Timokhin} A.~N.,  {Aharonian} F.~A.,   {Neronov} A.~Y.,  2004, \mn@doi [\aap]
  {10.1051/0004-6361:20040004}, \href
  {https://ui.adsabs.harvard.edu/abs/2004A&A...417..391T} {417, 391}

\bibitem[\protect\citeauthoryear{{V{\"o}lk}, {Aharonian}  \&
  {Breitschwerdt}}{{V{\"o}lk} et~al.}{1996}]{1996SSRv...75..279V}
{V{\"o}lk} H.~J.,  {Aharonian} F.~A.,   {Breitschwerdt} D.,  1996, \mn@doi
  [\ssr] {10.1007/BF00195040}, \href
  {https://ui.adsabs.harvard.edu/abs/1996SSRv...75..279V} {75, 279}

\bibitem[\protect\citeauthoryear{{Willson}}{{Willson}}{1970}]{radio1970MNRAS.151....1W}
{Willson} M.~A.~G.,  1970, \mn@doi [\mnras] {10.1093/mnras/151.1.1}, \href
  {https://ui.adsabs.harvard.edu/abs/1970MNRAS.151....1W} {151, 1}

\bibitem[\protect\citeauthoryear{{Wood}, {Caputo}, {Charles}, {Di Mauro},
  {Magill}, {Perkins}  \& {Fermi-LAT Collaboration}}{{Wood}
  et~al.}{2017}]{2017ICRC...35..824W}
{Wood} M.,  {Caputo} R.,  {Charles} E.,  {Di Mauro} M.,  {Magill} J.,
  {Perkins} J.~S.,   {Fermi-LAT Collaboration} 2017, in 35th International
  Cosmic Ray Conference (ICRC2017). p.~824 (\mn@eprint {arXiv} {1707.09551})

\bibitem[\protect\citeauthoryear{{Xi}, {Wang}, {Liang}, {Peng}, {Yang}  \&
  {Liu}}{{Xi} et~al.}{2018}]{fermicoma2018PhRvD..98f3006X}
{Xi} S.-Q.,  {Wang} X.-Y.,  {Liang} Y.-F.,  {Peng} F.-K.,  {Yang} R.-Z.,
  {Liu} R.-Y.,  2018, \mn@doi [\prd] {10.1103/PhysRevD.98.063006}, \href
  {https://ui.adsabs.harvard.edu/abs/2018PhRvD..98f3006X} {98, 063006}

\bibitem[\protect\citeauthoryear{{Zargaryan}, {Baghmanyan}, {Aharonian},
  {Yang}, {Casanova}  \& {Mackey}}{{Zargaryan} et~al.}{2021}]{zargaryan_coma}
{Zargaryan} D.,  {Baghmanyan} V.,  {Aharonian} F.,  {Yang} R.-Z.,  {Casanova}
  S.,   {Mackey} J.,  2021, POS, https://pos.sissa.it/395/582/pdf, \href
  {https://pos.sissa.it/395/582/pdf} {}

\bibitem[\protect\citeauthoryear{{Zhang}, {Churazov}  \& {Zhuravleva}}{{Zhang}
  et~al.}{2021}]{2021MNRAS.501.1038Z}
{Zhang} C.,  {Churazov} E.,   {Zhuravleva} I.,  2021, \mn@doi [\mnras]
  {10.1093/mnras/staa3718}, \href
  {https://ui.adsabs.harvard.edu/abs/2021MNRAS.501.1038Z} {501, 1038}

\bibitem[\protect\citeauthoryear{{van Weeren}, {de Gasperin}, {Akamatsu},
  {Br{\"u}ggen}, {Feretti}, {Kang}, {Stroe}  \& {Zandanel}}{{van Weeren}
  et~al.}{2019}]{radiohalo2019SSRv..215...16V}
{van Weeren} R.~J.,  {de Gasperin} F.,  {Akamatsu} H.,  {Br{\"u}ggen} M.,
  {Feretti} L.,  {Kang} H.,  {Stroe} A.,   {Zandanel} F.,  2019, \mn@doi [\ssr]
  {10.1007/s11214-019-0584-z}, \href
  {https://ui.adsabs.harvard.edu/abs/2019SSRv..215...16V} {215, 16}

\makeatother
\end{thebibliography}

\appendix

\section{Simulations}

To test the viability of the models considered in this work, we performed simulations for each one. To do this, first, we generated 1000 simulated datasets for each model using the best-fit parameters as input. It was done with \textit{simulate\_roi} tool in Fermipy, which takes the predicated number of counts for the given model and generates Poisson-distributed random (\textit{randomize=True}) data. Then, we run the fit on all simulated datasets, leaving free to vary the spectral parameters of all components in $\gamma$-ray excess models. We also left free the normalization of the diffuse background models and the sources within $5^{\circ}$ from the center of the ROI free. The distribution of spectral indices, energy fluxes, and the TS values for all models are displayed in Figure \ref{fig:fig3}.

\section{Systematic uncertainties}

Besides the simulations, we also estimated the main sources of systematic uncertainties on the single disk model according to \citep{fermicoma2018PhRvD..98f3006X}, namely the Galactic interstellar emission model (IEM), the selection of event class, and the choice of the low energy threshold. To estimate the systematic effects due to the IEM, we employed eight alternative components of the spectral line surveys of HI and IC emission used in \citet{2016ApJS..224....8A} in the estimation of systematic uncertainties on the fitted properties of SNRs. For the diffuse isotropic emission, we used the same template fixing the normalization to the best-fit value found in the main analysis varying it by $\mathrm{(\pm2\sigma)}$ compatible with the statistical errors of the isotropic component of the main analysis. For the event class selection, first, we applied the disk model on the baseline models created with SOURCE and CLEAN classes, which have larger effective areas at the expense of higher contamination of background events. And finally, we performed the spectral analysis with the disk model by varying the energy threshold between 100 MeV and 500 MeV. The combined systematic errors on the best-fitted spectral parameters are summarized in Table \ref{tab:table3}.

\label{lastpage}
\end{document}